# Immunohistochemical pitfalls in the demonstration of insulin-degrading enzyme in normal and neoplastic human tissues

RAZVAN T. RADULESCU[1*], ANGELIKA JAHN[1], DANIELA HELLMANN[1], AND GREGOR WEIRICH[2]

[1]*Clinical Research Unit, Department of Obstetrics and Gynecology, Klinikum rechts der Isar, Technical University of Munich, Ismaninger Str. 22, 81675 Munich, Germany*

[2]*Institute of Pathology, Technical University of Munich, Ismaninger Str. 22, 81675 Munich, Germany*

*Corresponding author. Present affiliation: *c/o Munich Agency for Work, 80337 Munich*.
E-mail: artiaris@yahoo.de

## ABSTRACT

Previously, we have identified the cytoplasmic zinc metalloprotease insulin-degrading enzyme (IDE) in human tissues by an immunohistochemical method involving no antigen retrieval (AR) by pressure cooking to avoid artifacts by endogenous biotin exposure and a detection kit based on the **l**abeled **s**trept**a**vidin **b**iotin (LSAB) method. Thereby, we also employed 3% hydrogen peroxide ($H_2O_2$) for the inhibition of endogenous peroxidase activity and incubated the tissue sections with the biotinylated secondary antibody at room temperature (RT). We now add the immunohistochemical details that had led us to this optimized procedure as they also bear a more general relevance when demonstrating intracellular tissue antigens. Our most important result is that endogenous peroxidase inhibition by 0.3% $H_2O_2$ coincided with an apparently positive IDE staining in an investigated breast cancer specimen whereas combining a block by 3% $H_2O_2$ with an incubation of the biotinylated secondary antibody at RT, yet not at $37^0C$, revealed this specimen as almost entirely IDE-negative. Our present data caution against three different immunohistochemical pitfalls that might cause falsely positive results and artifacts when using an LSAB- and peroxidase-based detection method: pressure cooking for AR, insufficient quenching of endogenous peroxidases and heating of tissue sections while incubating with biotinylated secondary antibodies.

**Introduction**

Immunohistochemistry (IHC) is a widely used approach to visualize antigens in various cells including those of primary tissues taken from patients. Meanwhile, well-defined procedures have evolved that keep potential errors to a minimum. Among the most common pitfalls are endogenous biotin and endogenous peroxidases. The peril of endogenous biotin confounding specific staining signals has previously been addressed (Bussolati *et al.* 1997, Iezzoni *et al.* 1999, Kim *et al.* 2002).

As part of our staining for insulin-degrading enzyme (IDE) in normal and pathologic human tissues by using a widely used detection technique abbreviated LSAB, i.e. the **l**abeled **s**trept**a**vidin **b**iotin method (Radulescu *et al.*, 2007), we have successfully minimized the exposure of endogenous biotin molecules by avoiding a pressure cooking-based antigen retrieval (AR) since, among several techniques compared to one another, this type of AR has been shown to maximally expose endogenous biotin (Kim *et al.* 2002).

In the course of closing in on the optimal IHC method for demonstrating IDE, we have also considered two further important potential causes for falsely positive staining when using the LSAB technique: endogenous peroxidases and heating of tissue sections during the incubation with biotinylated secondary antibodies.

The distribution of endogenous peroxidases is as ubiquitous as that of endogenous biotin. Among the human tissues abundant in peroxidases are normal kidney (Chu *et al.* 1992) as well as normal breast (Chu *et al.* 1992) and breast carcinomas (di Ilio *et al.* 1985). In order to improve the signal-to-noise ratio in IHC staining of such non-hematologic tissues, the unspecific background peroxidase activity needs to be blocked (van Bogaert *et al.* 1980). This is frequently achieved by incubating the formalin-fixed tissue sections with a solution of 3% hydrogen peroxide ($H_2O_2$) in water (Boenisch 2003) or in methanol (Noll *et al.* 2000) prior to the application of the primary antibody.

We present here our comparison of two different peroxidase blockades by means of solutions of 3% vs. 0.3% $H_2O_2$ dissolved in water. In the same context, we have also investigated the influence of the incubation temperature of the biotinylated secondary antibody on the staining intensity of the IDE antigen as another possible source of artifacts.

**Materials and methods**

Formalin-fixed, paraffin-embedded specimens of human tissues- normal kidney, non-malignant mammary gland and breast cancer- were drawn at random from the archives of the Institute of Pathology of the Technical University of Munich (Germany). Immunohistochemical staining for IDE was performed as follows: First, the sections were deparaffinized and hydrated by passing them through xylene twice for 10 min and a graded series of ethanol going from 100% ethanol twice for 5 min to 96% ethanol once for 5 min and finally to 70% ethanol for 5 min. After two 2.5-min washing steps in tris-buffered saline (TBS) buffer at room temperature (RT), antigen retrieval (AR) was performed by boiling the slides in 10 mmol/l citrate (Sigma C-1909, MW 210.1) buffer (pH 6.0) for 4 min in order to also assess the amount of endogenous biotin released by this procedure, as previously described (Kim *et al.* 2002), this AR being ultimately dismissed as a result of this series of experiments (cf. below). After two 2.5-min washing steps in TBS at room temperature (RT), endogenous peroxidase activity was quenched by incubating the slides either with a 0.3% or a 3% hydrogen peroxide ($H_2O_2$) solution- prepared from a 30% stock solution (K33354110, 1.07210.0250, Merck, Darmstadt, Germany) by a 1:100 or, respectively, a 1:10 dilution in distilled water- at RT for 20 min. Subsequent to two 2.5-min washing steps in TBS at RT, a protein block with a 10% solution of normal goat serum (X0907, Dako) was performed for 30 min and, after decanting this serum, the tissue sections were incubated at $4°C$ overnight with UCG 43/6, a rabbit polyclonal antibody to recombinant full-length human IDE (Chesneau *et al.* 2000), that had been diluted 1:1000 in antibody diluent (S2022, Dako, Hamburg, Germany). According to the manufacturer, this diluent has the property of reducing non-specific antibody binding. The procedure resumed the next day with two 2.5-min washing steps in TBS at RT. Subsequently, a biotin-streptavidin-peroxidase-based detection kit (K5003, Dako, Hamburg, Germany) was used whereby the biotinylated goat anti-rabbit secondary antibody was either: a) incubated at $37°C$ for 30 min followed by a 5-min washing step and then a 30-min incubation with streptavidin-peroxidase at RT; or b) incubated at RT for 30 min followed by a 5-min washing step and then a 30-min incubation with streptavidin-peroxidase at RT. After another identical washing step (cf. above), the chromogenic reaction was carried out with the peroxidase substrate AEC (K3464, Dako, Hamburg, Germany) at RT for 10 min. After a final washing step, nuclei were counterstained in Mayer´s acid hematoxylin for 10 sec. Subsequently, the slides were rinsed under running tap water, transferred to distilled water and mounted with Kaiser´s glycerol gelatin (Merck, Darmstadt, Germany).

**Results**

By employing a rabbit polyclonal antibody to recombinant human IDE (Chesneau *et al.* 2000) within an immunohistochemical procedure involving pressure cooking for antigen retrieval (AR), a block of endogenous peroxidases by means of 0.3% hydrogen peroxide ($H_2O_2$) and an incubation of the tissue sections with biotinylated secondary antibody at $37^0C$, we obtained apparent staining for IDE in normal kidney (Fig. 1a), non-malignant mammary gland (Fig. 1c) and breast cancer (Fig. 1e). Notably, the non-malignant mammary gland duct epithelial cells were homogenously positive (Fig. 1c). Moreover, while the corresponding controls performed without the addition of the antibody to IDE were negative for the non-malignant (Fig. 1d) and malignant (1f) breast tissue sections, the respective kidney section (Fig. 1b) was, however, positive, specifically and as expected, the renal tubule epithelial cells, yet not the glomeruli.

We also performed staining for IDE involving the use of a 3% $H_2O_2$ solution while maintaining the parameter of the incubation of the biotinylated secondary antibody at $37^0C$ (Fig. 2). This increase in the strength of the endogenous peroxidase block did not impair IDE staining in the normal kidney tissue section (Fig. 2a), yet it could not prevent that the kidney control designed to be negative was again positive (Fig. 2b). Also under these conditions, the non-malignant mammary gland duct epithelial cells were homogenously positive (Fig. 2c). Apparent staining for IDE was also obtained in the breast cancer tissue section (Fig. 2e). The negative controls for the non-malignant mammary gland (Fig. 2d) and breast cancer (Fig. 2e) specimens were negative.

Lowering the temperature at which the tissue sections were incubated with the biotinylated secondary antibody from $37^0C$ to RT, however, yielded an interesting IDE staining difference in the non-malignant mammary gland specimen whereas IDE staining in the normal kidney remained largely unaffected by this variation. Accordingly, while the normal kidney tissue sections stained positively for IDE (Fig. 3a and Fig. 4a), such ambient temperature conditions, combined with either a 0.3% or 3% $H_2O_2$ block of endogenous peroxidase activity, led to a *heterogenous* IDE staining pattern in the non-malignant mammary gland (Fig. 3c and Fig. 4c, respectively), thus contrasting with the homogenous picture of the non-malignant human breast obtained when the tissue sections were heated at $37^0C$ during the incubation with the biotinylated secondary antibody (cf. above).

The corresponding controls performed without the addition of the antibody to IDE were negative for the non-malignant (Fig. 3d and Fig. 4d, respectively) and malignant (Fig. 3f and Fig. 4f, respectively) breast tissue sections, yet the respective normal kidney tissue sections were again positive (Fig. 3b and Fig. 4b).

We also noted that our procedural variations had the most dramatic consequences with regard to the pattern of staining in the breast carcinoma sections. As such, we could ascertain a clear reduction in staining intensity when increasing the strength of the endogenous peroxidase block (Fig. 2e vs. Fig. 1e) or when decreasing the temperature of incubation with the biotinylated secondary antibody (Fig. 3e vs. Fig. 1e). By combining these optimized conditions, i.e. the endogenous peroxidase block with a 3% $H_2O_2$ solution and the incubation of the biotinylated secondary antibody at RT (Fig. 4e), the same breast carcinoma tissue section that had appeared as IDE-positive with a less concentrated $H_2O_2$ solution and during heat incubation with the biotinylated secondary antibody (Fig. 1e), was in fact almost entirely IDE-negative.

To even better illustrate this differential picture in the investigated breast carcinoma section, a direct juxtaposition of the four different procedural conditions is shown in Fig. 5.

Finally, the unspecific staining in the negative kidney controls was only abolished by avoiding pressure cooking for AR, employing a 3% $H_2O_2$ solution for the inhibition of endogenous peroxidases and incubating the biotinylated secondary antibody at RT (Fig. 6).

**Discussion**

A recent report has thoroughly demonstrated for a number of normal human tissues the importance of endogenous biotin as a potential pitfall in immunohistochemistry, mainly by means of an antibody specific for biotin (Kim *et al.* 2002). While these authors compared procedures that differed in the method of antigen retrieval (AR) and thereby found that pressure cooking maximally exposes endogenous biotin, their block of endogenous peroxidases was constantly performed by means of 3% hydrogen peroxide ($H_2O_2$).

In our recent immunohistochemical analysis for IDE in human tissues by means of the LSAB detection technique (Radulescu *et al.*, 2007), we have considered both of these aspects by avoiding pressure cooking for AR and concomitantly using a 3% $H_2O_2$ solution for the inhibition of intrinsic tissue peroxidase activity. Despite these measures, IDE staining, albeit with a five-fold higher primary antibody concentration than used here, was not compromised (Radulescu *et al.*, 2007). We report here further important details that have led us to this optimized immunohistochemical procedure.

As such, we have compared the effectiveness of 3% $H_2O_2$ vs. 0.3% $H_2O_2$ in quenching endogenous peroxidases and, moreover, the effect of incubating the biotinylated secondary antibody at $37^0C$ vs. RT. From the investigated breast carcinoma sections it became particularly obvious that staining artifacts are caused by either an insufficient endogenous peroxidase block with 0.3% $H_2O_2$ or a heating at $37^0C$ when incubating with the biotinylated secondary antibody. These falsely positive immunhistochemical reactions could be largely abolished only when both of these conditions were optimized, specifically by employing a procedure in which a 3% $H_2O_2$ solution was used for the block of endogenous peroxidases and the biotinylated secondary antibody was incubated at RT.

The importance of a reference tissue both as a positive and negative control should also be emphasized, e.g. the relevance of assessing normal kidney positive and negative controls when evaluating the IDE staining results in normal or neoplastic human breast tissues. As such, we have conclusively revealed by means of the normal human kidney negative controls that there is a significant unspecific contribution of endogenous biotin to the IDE staining result on those tissue sections to which the antibody to IDE had been added.

The validity of this point is reinforced by the immunohistochemical constellation described by Iezzoni *et al.* for the tissue antigen inhibin in human tissues whereby tissues unblocked for endogenous biotin stained falsely positive for inhibin, yet upon appropriate endogenous biotin inhibition turned out to be inhibin-negative (Iezzoni *et al.* 1999).

Therefore, the most important lesson from our investigations has been that normal and neoplastic breast positive and negative controls are necessary but *not* sufficient whereas normal kidney positive and negative controls are both necessary *and* sufficient as controls. In this context, our present data have relevance not only for peroxidase-rich tissues such as breast cancer, but also for those tissues with a relatively lower activity of peroxidases such as normal breast or normal tissue sections adjacent to mammary carcinoma cells (Kumaraguruparan *et al.* 2002).

The reason for this is that in such tissues or tissue areas, the negative control- achieved by leaving out the primary antibody to a *non*-peroxidase antigen (e.g. IDE)- may be negative even if peroxidase blockade is suboptimal or insufficient, yet it may simply reflect a light microscopic detection limit for the complexes between endogenous peroxidase and its chromogen substrate added at the end of the staining procedure.

In contrast, an *antibody specifically recognizing an endogenous peroxidase* would clearly reveal the presence of such a peroxidase even in such tissues, similar to what has previously been shown by use of an antibody specific for biotin in various human tissues (Kim *et al.* 2002). Consistent with this assumption, it has been shown that many normal human tissues including breast and kidney do indeed express the gene for glutathione peroxidase which is one of the major endogenous peroxidases (Chu *et al.* 1992).

Therefore, when interpreting a peroxidase-based staining result in those sections to which the primary antibody specific for the *non*-peroxidase antigen has been applied, the key point is that the perceived color intensity may represent an *additive contribution* from insufficiently blocked endogenous peroxidases and the added streptavidin-peroxidase component of the detection kit.

As a result, when it ultimately comes to scoring the staining intensities of normal vs. diseased tissues or yet between two different kinds of diseased tissues for diagnostic and/or prognostic purposes, such considerations become particularly critical. For instance, a tissue which stains negatively for a given protein of interest after appropriate endogenous peroxidase blockade, may, by contrast, score 1+, i.e. positive, albeit weakly positive, if the block is inadequate and, hence, this tissue would be falsely positive for that protein. In order to avoid such misleading *overstaining* or even *falsely positive* staining, one has to block endogenous peroxidase activity in an appropriate manner, i.e. with 3% $H_2O_2$.

As we have also shown here, the incubation of the biotinylated secondary antibody should be performed at room temperature, and not at $37^0C$, in order to avoid artifactual overstaining or yet falsely positive staining when employing the LSAB detection technique.

Taken together, our present data on potential pitfalls in the immunohistochemistry of IDE should further increase the awareness for avoiding the exposure of endogenous biotin as well as for an adequate block of endogenous peroxidases and the right incubation temperature when staining a given antigen. Ultimately, due consideration of these aspects should increase the comparability of different immunohistochemical studies on the same antigen in various tissue specimens and, moreover, contribute to avoiding erroneous interpretations with otherwise potentially detrimental consequences for the diagnosis, treatment and prognosis of human diseases.

**Acknowledgements**

We thank Marsha Rosner for providing the antibody to insulin-degrading enzyme, Manfred Schmitt for helpful comments and Tom Boenisch for positive feedback. This work was supported in part by a grant from the Federal Institute for Drugs and Medical Devices, Bonn, Germany. The contents and form of the present manuscript are essentially identical to those of its initial pre-publication version that was accomplished on December 7, 2006.

**Figure legends**

**Figure 1**   Immunohistochemical demonstration of IDE by means of a rabbit polyclonal antibody to IDE in formalin-fixed, paraffin-embedded normal human kidney (A), non-malignant human breast (C) and human breast carcinomas (E) whereby tissue sections were subjected to pressure cooking for antigen retrieval, their endogenous peroxidases blocked with 0.3% $H_2O_2$ and the biotinylated secondary antibody included in DAKO detection kit K5003 was added onto the sections at $37^0$C. The corresponding negative controls (B, D, F) were performed without addition of the primary antibody to IDE. Objective x20.

**Figure 2**   Immunohistochemical demonstration of IDE by means of a rabbit polyclonal antibody to IDE in formalin-fixed, paraffin-embedded normal human kidney (A), non-malignant human breast (C) and human breast carcinomas (E) whereby tissue sections were subjected to pressure cooking for antigen retrieval, their endogenous peroxidases blocked with 3% $H_2O_2$ and the biotinylated secondary antibody included in DAKO detection kit K5003 was added onto the sections at $37^0$C. The corresponding negative controls (B, D, F) were performed without addition of the primary antibody to IDE. Objective x20.

**Figure 3**   Immunohistochemical demonstration of IDE by means of a rabbit polyclonal antibody to IDE in formalin-fixed, paraffin-embedded normal human kidney (A), non-malignant human breast (C) and human breast carcinomas (E) whereby tissue sections were subjected to pressure cooking for antigen retrieval, their endogenous peroxidases blocked with 0.3% $H_2O_2$ and the biotinylated secondary antibody included in DAKO detection kit K5003 was added onto the sections at RT. The corresponding negative controls (B, D, F) were performed without addition of the primary antibody to IDE. Objective x20.

**Figure 4**   Immunohistochemical demonstration of IDE by means of a rabbit polyclonal antibody to IDE in formalin-fixed, paraffin-embedded normal human kidney (A), non-malignant human breast (C) and human breast carcinomas (E) whereby tissue sections were subjected to pressure cooking for antigen retrieval, their endogenous

peroxidases blocked with 3% $H_2O_2$ and the biotinylated secondary antibody included in DAKO detection kit K5003 was added onto the sections at RT. The corresponding negative controls (B, D, F) were performed without addition of the primary antibody to IDE. Objective x20.

**Figure 5** Immunohistochemical demonstration of IDE by means of a rabbit polyclonal antibody to IDE in formalin-fixed, paraffin-embedded human breast carcinomas whereby tissue sections were subjected to pressure cooking for antigen retrieval, their endogenous peroxidases blocked with 0.3% $H_2O_2$ (A,C) or 3% $H_2O_2$ (B,D) and the biotinylated secondary antibody included in DAKO detection kit K5003 was added onto the sections at $37^0C$ (A,B) or at RT (C,D). Objective x20.

**Figure 6** Formalin-fixed, paraffin-embedded normal human kidney section whereby the primary antibody to IDE has been left out during the immunohistochemical procedure. No antigen retrieval (neither pressure cooking nor any chemical method) was performed. Endogenous peroxidases were blocked with 3% $H_2O_2$. The biotinylated secondary antibody included in DAKO detection kit K5003 was incubated at RT. Objective x20.

**Figure 1**

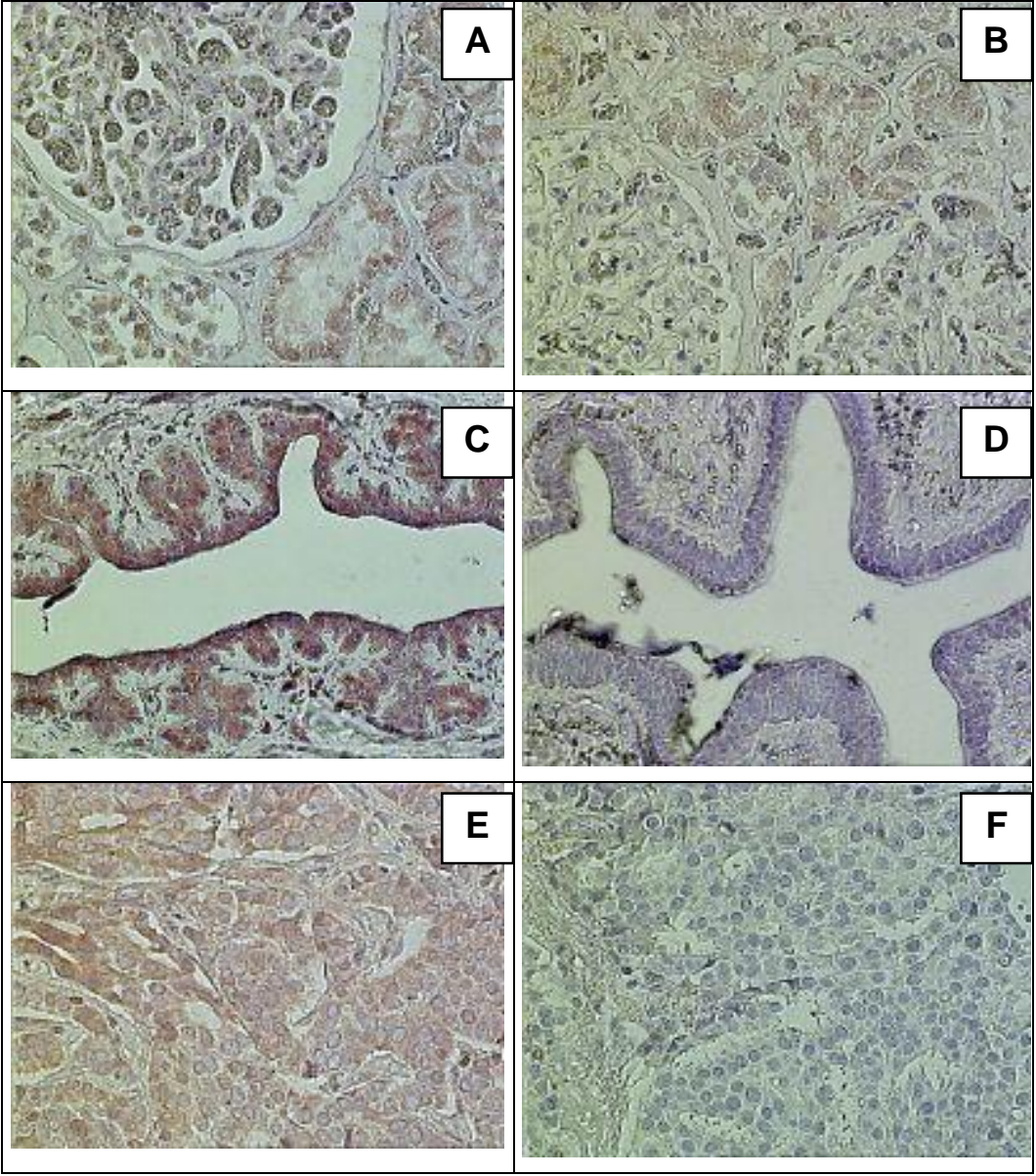

**Figure 2**

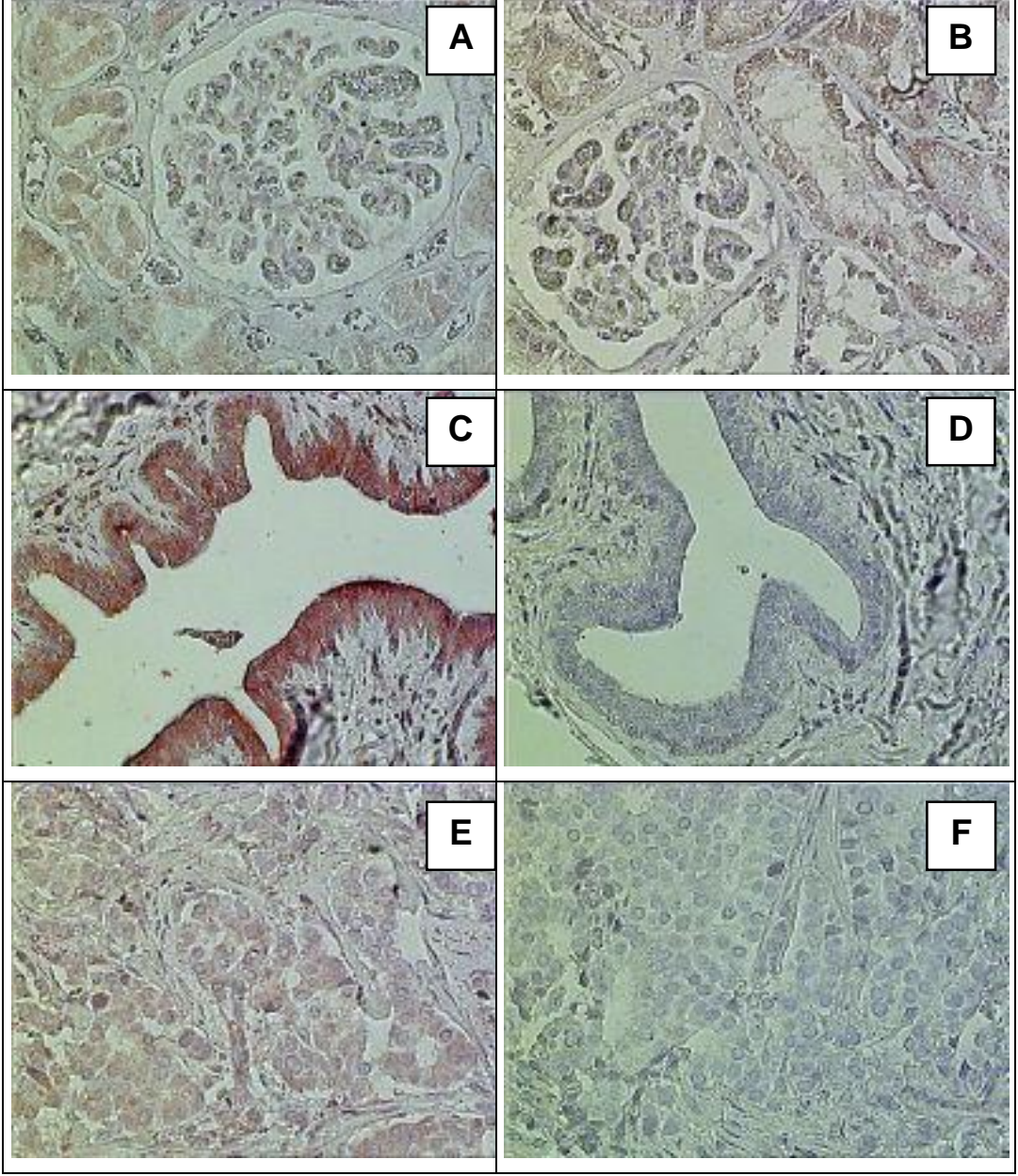

**Figure 3**

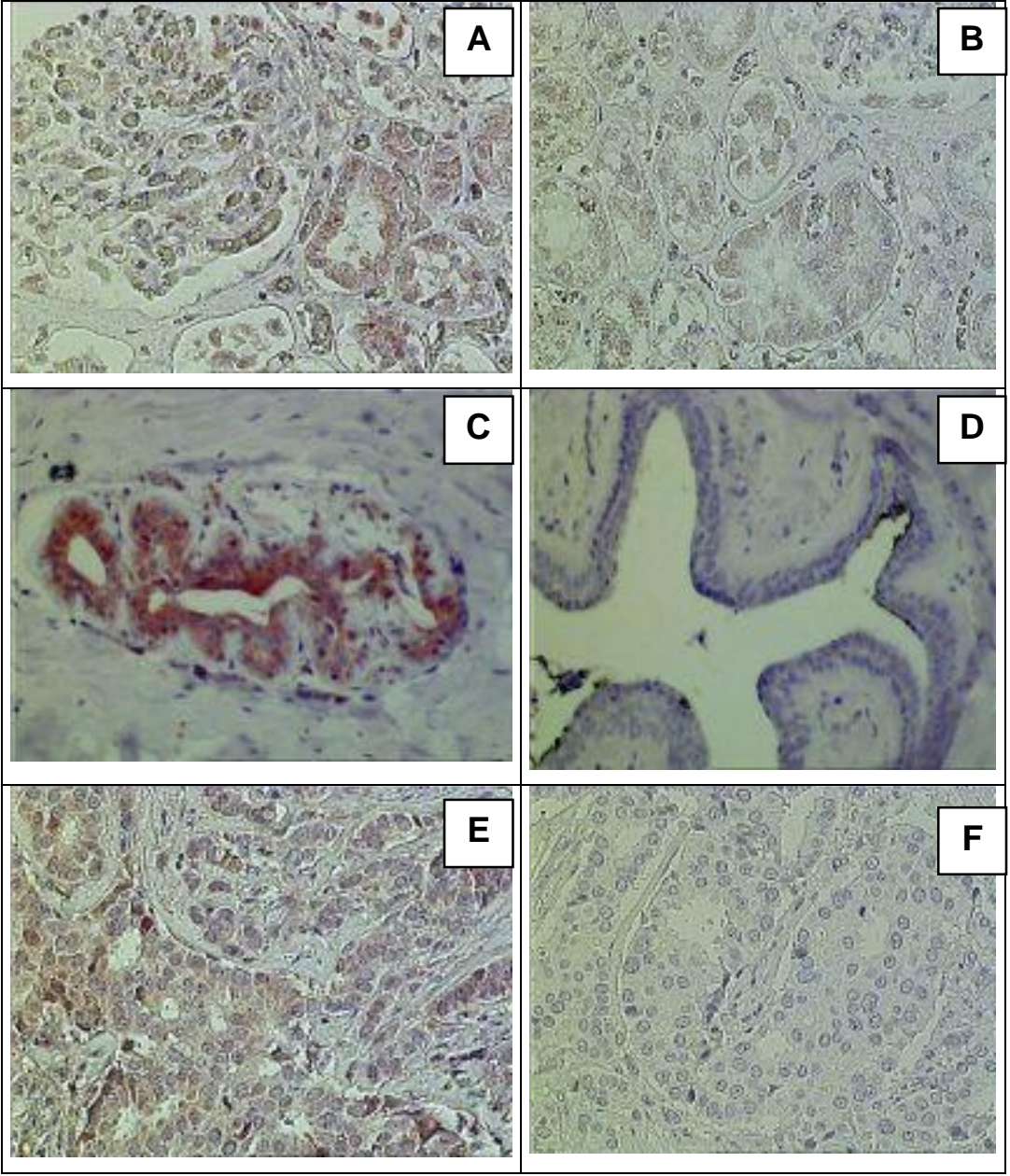

**Figure 4**

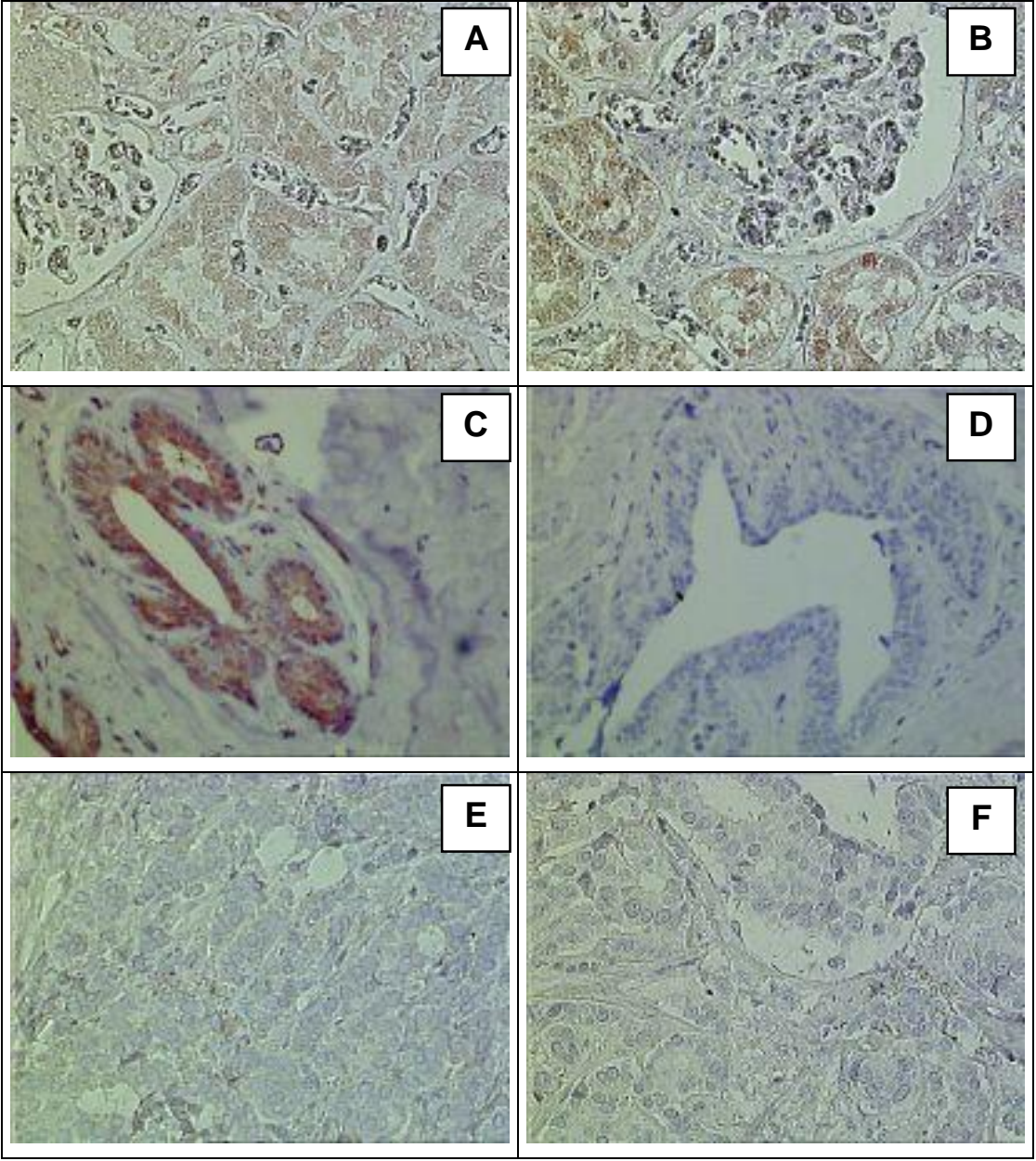

**Figure 5**

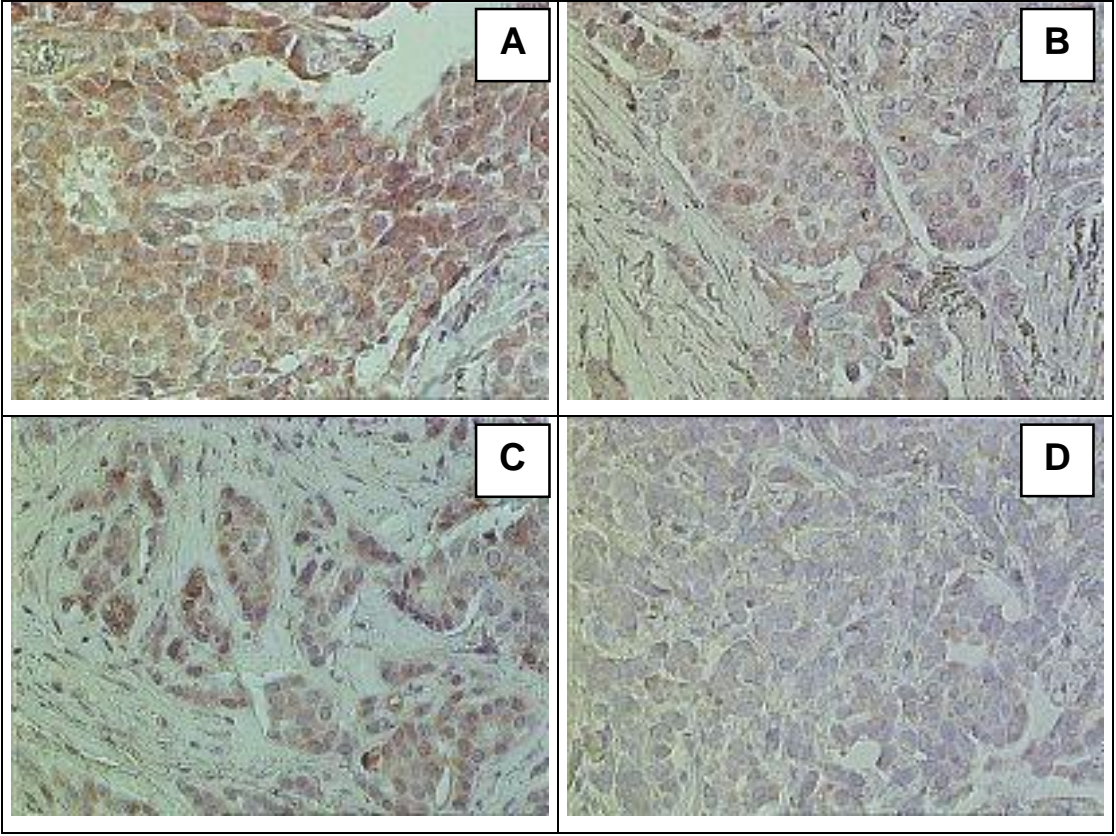

**Figure 6**

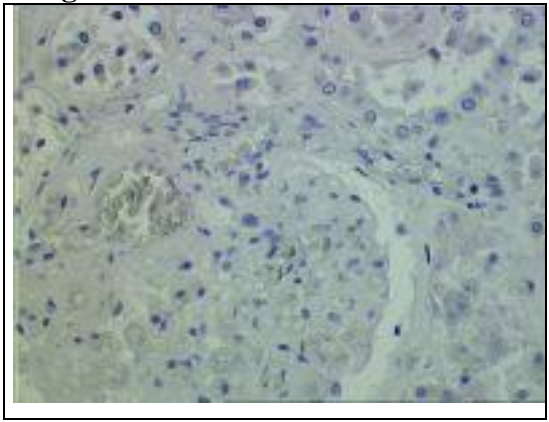